\documentclass[]{aa}  

\usepackage{graphicx}
\usepackage{txfonts}
\usepackage{xspace}

\newcommand{\Ha}{H$\alpha$\xspace}

\newcommand{\highelf}{{\sc highelf}\xspace}
\newcommand{\sinopsis}{{\sc sinopsis}\xspace}
\newcommand{\ma}{$\rm M_\ast$\xspace}

\newcommand{\Ssfr}{$\rm \Sigma_{SFR}$\xspace}
\newcommand{\Sm}{$\rm \Sigma_\ast$\xspace}

\newcommand{\msk}{$\rm M_\odot \, kpc^{-2}$\xspace}

\begin{document}

   \title{Evidence for enhanced star formation rates in $z\sim 0.35$ cluster galaxies undergoing ram pressure stripping}

 \titlerunning{SFR-Mass relation in RPS galaxies at z$\sim$0.35}
 
   \author{Benedetta Vulcani\inst{1}
          \and Alessia Moretti\inst{1}
          \and Bianca M. Poggianti\inst{1}
          \and Mario Radovich\inst{1}
          \and Ariel Werle\inst{1}
          \and Marco Gullieuszik\inst{1}
          \and Jacopo Fritz\inst{2}
          \and Cecilia Bacchini\inst{1}
           \and Johan Richard\inst{3}
          }

   \institute{INAF- Osservatorio astronomico di Padova, Vicolo Osservatorio 5, I-35122 Padova, Italy\\
              \email{benedetta.vulcani@inaf.it}
         \and
             Instituto de Radioastronomia y Astrofisica, UNAM, Campus Morelia, AP 3-72, CP 58089, Mexico
         \and
            Univ. Lyon, Univ Lyon1, ENS de Lyon, CNRS, Centre de Recherche Astrophysique de Lyon UMR5574, F-69230 Saint-Genis-Laval, France
                          }

 \authorrunning{Vulcani et al. }
   \date{}

 
  \abstract
{Ram pressure stripping (RPS) is one of the most invoked mechanisms to explain the observed differences between cluster and field galaxies. In the local Universe, its effect on the galaxy star forming properties has been largely tackled and the general consensus is that this process first compresses the gas available in the galaxy disks, boosting the star formation  for a limited amount of time, and then removes the remaining gas leading to quenching. Much less is  known on the effect and preponderance of RPS at higher redshift, due to the lack of statistical samples. Exploiting VLT/MUSE observations of  galaxies at $0.2<z<0.55$ and the catalog of ram pressure stripped galaxies by Moretti et al., we  compare the global star formation rate–mass  (SFR-\ma) relation of 29 cluster galaxies undergoing RPS to that of 26 field and cluster undisturbed galaxies that consitute our control sample. Stripping galaxies occupy the upper envelope of the control sample SFR–\ma relation, showing a systematic enhancement of the SFR at any given mass. The boost is $>3\sigma$ when considering the SFR occurring in both the tail and disk of galaxies. The enhancement is retrieved also on local scales: considering spatially resolved data, ram pressure stripped galaxies overall have large \Ssfr values, especially for \Sm$>10^{7.5}$\msk.
RPS seems to leave the same imprint on the SFR-\ma and \Ssfr-\Sm relations both in the Local Universe and at $z\sim 0.35$.}


   \maketitle
%

\section{Introduction} \label{sec:intro}

Multiple observations support the evidence that galaxies residing in dense cluster environments undergo distinct evolutionary pathways compared to their counterparts in less crowded regions of the universe. 
For example, group and cluster galaxies are HI deficient \citep{GiovanelliHaynes1985}, have redder colors \citep{Kennicutt1983, Kodama2001} and a lower star formation rate \citep[SFR,][]{BowerBalogh2004, Vulcani2010, Perez2023}, and exhibit non-disky  morphologies more frequently \citep{Dressler1980, Vulcani2023} than similar-mass analogues in the field. Such differences arise from the fact that as  galaxies navigate the high-density environments of clusters or groups, their properties can be profoundly impacted by various physical processes, such as tidal interactions \citep{Moore1996}, starvation or strangulation \cite{Larson1980}, ram pressure stripping \citep[RPS,][]{Gunn1972}, and, to less extent, galaxy-galaxy interactions. These mechanisms can alter the availability of cold gas within galaxies, which in turn influences their ability to form stars. In particular, RPS has been shown to be capable of pulling out of  galaxies not only the loosely-bound circumgalactic medium (CGM), but also the cold and dense interstellar medium (ISM), producing tails of gas in the direction opposite to a galaxy’s direction of motion \citep[e.g.,][]{Cortese2007, Yagi2007, Sun2010, Fumagalli2014, Poggianti2017GASPMuse, Gullieuszik2017, Bellhouse2017, Jachym2017, Cramer2019, Moretti2020, Bacchini2023}. 

Since RPS removes the  interstellar gas reservoirs, its long term impact is to suppress and even halt star formation in cluster galaxies \citep{Vollmer2001, Tonnesen2007, Vulcani2020}. Nonetheless, multiple observations have shown that, at least in the Local Universe, galaxies undergoing RPS are having a burst in star formation in their disks due to gas compression \citep{Poggianti2016JELLYFISHREDSHIFT, Vulcani2018_L, RobertsParker2020, Roberts2022b, Boselli2023} or to  ram pressure-driven mass flows \citep{Zhu2023},  and star formation can be induced in the tails as well \citep{Vulcani2018_L, Cramer2019, Jachym2019, Gullieuszik2020, Poggianti2019}. As a result, according to these observations ram pressure stripped galaxies are characterized by globally-enhanced SFRs \citep{Vulcani2018_L, Ramatsoku2020, Vulcani2020b, RobertsParker2020} in comparison to normal star-forming galaxies in clusters and in the field. 
Nonetheless, other observations have shown no signs of enhanced SFR or have even found it reduced in ram pressure stripped galaxies \citep{Yoon2017, Mun2021}, leaving the  question of the actual role of  RPS on SFRs  open.

From the theoretical point of view, many efforts have been made  to simulate ram pressure stripped galaxies, with a variety of codes, techniques, and included physical processes. 
Different simulation approaches have given different outcomes. Simulations relying on wind-tunnel setups,  overall find that RPS boosts the SFR \citep{Kronberger2008, Kapferer2009,TonnesenBryan2012, Roediger2014}. 

{ Exploring hydrodynamical simulations,} 
\cite{Steinhauser2016} found a general enhancement of SFR only for galaxies experiencing mild ram pressure; \cite{RamosMartinez2018} 
showed that magnetic fields can channel gas to the centre of the galaxy where it can be a reservoir for star formation.

Exploiting full cosmological galaxy simulations, \cite{Troncoso2016, Troncoso2020} showed that the EAGLE \citep{Schaye2015} simulation predicts an enhancement of star formation in the so-called leading half of a galaxy falling into a cluster, whereas in the trailing half no increase in SFR is found. In contrast, TNG50 simulation \citep{Pillepich2019, Nelson2019} does not predict higher SFRs in ram pressure stripped galaxies compared to analog cluster galaxies with the same stellar mass or gas fraction \citep{2023arXiv230409199G}. Nonetheless, it does predict both star formation within the ram pressure-stripped tails and  bursts of elevated star formation along the history of stripped galaxies, even though these do not impact the global SFR values  \citep{2023arXiv230409199G}. 

All the studies cited above have focused on characterizing the consequences of stripping on the galaxy life cycle  in the Local Universe, where most of the systematic observational efforts have been conducted (e.g. GASP,  \citealt{Poggianti2017GASPMuse}; VIVA, \citealt{Chung2009};  LoTSS, \citealt{Roberts2021a}).  Much less is instead known at $z\geq0.1$. 
{ Numerical simulations \citep{Singh2019} and analytic prescriptions \citep{Fujita2001} predict an amplification of RPS efficacy as redshift increases, due to the dependence of RPS on the ICM density \citep{KravtsovBorgani2012, Mostoghiu2019}. Observationally,  }
only a limited number of systematic searches have been executed \citep[e.g.,][]{Cortese2007}. \cite{Owers2012, Ebeling2014, Rawle2014, McPartland2016,  Durret2021}  exploited {\it HST} imaging and identified ram pressure stripping candidates characterized by star-forming tails emitting in the bluer  {\it Hubble Space Telescope (HST)} bands at $0.3<z<0.7$. Lacking spectroscopic confirmation in most of the cases (except for \citealt{Owers2012,  Rawle2014}), though, these galaxies remain candidates and are not confirmed to be undergoing ram pressure stripping. The first detailed characterization of a ram pressure stripped galaxy at $z\sim0.7$ has been possible thanks to the advent of the Multi Unit Spectroscopic Explorer at the Very Large Telescope (MUSE/VLT) Integral field spectrograph \citep{Boselli2019}.
Exploiting the MUSE Guaranteed Time Observations (GTO) data \citep{Bacon2017, Richard2021}, \cite{Moretti2022} and A. Moretti et al. (in prep.), assembled the first, and up to now only, large sample of spectroscopically confirmed ram pressure stripped galaxies at intermediate redshift ($z\sim$0.3–0.5). It is now possible to investigate the properties of ionized gas tails due to ongoing RPS beyond the local universe, to understand whether the effect of this process on galaxies varies with time. 

In this paper we exploit the A. Moretti et al. (in prep.) catalog and  delve into the multifaceted nature of the star formation rate - stellar mass (SFR-\ma) relation for cluster galaxies undergoing RPS at z$\sim 0.35$. While in general a well-defined correlation between star formation rate and mass exist \citep{Noeske2007a, Noeske2007b}, with more massive galaxies exhibiting higher rates of star formation due to their larger gas reservoirs; the many cluster specific mechanisms, and in particular RPS, can have profound effects on the galaxy's ability to form new stars and, consequently, impact its position on the SFR-\ma relation. Exploiting the GASP sample, 
\cite{Vulcani2018_L} found that stripping galaxies occupy the upper envelope of a control sample (made of undisturbed galaxies in different environments) SFR-\ma relation, showing a systematic enhancement of the SFR at any given stellar mass. 
Exploiting spatially resolved data to investigate the origin and location of the excess, \cite{Vulcani2020b} found that even on $\sim$1kpc scales, stripping galaxies present a systematic enhancement of SFR density ($\Sigma$ SFR)  at any given mass density ($\Sigma_\ast$) compared to their undisturbed counterparts. This enhancement is proportional to the global SFR enhancement, for both stripped and non-stripped galaxies. 

In this paper, we aim at establishing if RPS is able to significantly affect the star forming properties of galaxies also at $z\sim 0.35$, as it does in the Local Universe, for the first time on a statistically significant sample. An attempt was already performed by \cite{Lee2022}, who, using Gemini GMOS/IFU observations for only 5 ram pressure stripped galaxies at $z\sim 0.3-0.4$, indeed found hints for an increase of star formation  compared to undisturbed galaxies. 

The paper is organized as follows: \S\ref{sec:data} presents the data sample and the sample selection. \S\ref{sec:analysis} summarizes the techniques adopted to extract the quantities of interest for our analysis, \S\ref{sec:results} presents the results of the analysis. We will characterize both the global and spatially resolved SFR-\ma relations. Finally, \S\ref{sec:summary} will summarize and discuss the results. 

Throughout the paper,  we adopt a \cite{Chabrier2003} initial mass function (IMF) in the mass range 0.1-100 M$_{\odot}$. The cosmological constants assumed are $\Omega_m=0.3$, $\Omega_{\Lambda}=0.7$ and H$_0=70$ km s$^{-1}$ Mpc$^{-1}$. 

\section{Data Sample and Sample selection} \label{sec:data}

We exploit MUSE observations gathered in the context of the MUSE GTO \citep{Richard2021}, which observed a set of clusters extracted from the Massive Clusters Survey \citep{Ebeling2001}, Frontier Fields \citep{Lotz2017}, GLASS \citep{Treu2015} and Cluster Lensing and Supernova survey with Hubble \citep{Postman2012} programmes. Clusters were observed with single pointings or mosaics, with exposure times ranging from $\sim$2 to $\sim$15 hr (effective). Observations were limited to only the central regions of the clusters: the radius of the typical area  covered by MUSE observations corresponds to $\sim$250–330 kpc, depending on the cluster's redshift, roughly matching the inner $\sim$0.1–0.15 R$_{200}$. The 5$\sigma$ emission-line detection limit for a point-like source  is in the range between (0.77-1.5)$\rm \sim 10^{-18} erg/s/cm^2$ at 7000 \AA{}. More details on the observations and data analysis can be found in \cite{Richard2021}.

Given the seeing conditions of the MUSE observations, we can characterize galaxy properties on a scale of 4-6 kpc/\arcsec, depending on redshift. 

As described in \cite{Moretti2022, Werle2022}, we selected 12 clusters in the redshift range 0.3$<$$z$$<$0.5 and identified galaxies 1) in the clusters and undergoing RPS; 2)  in the clusters or coeval field appearing undisturbed; 3) passive galaxies. Three of us (B.P., B.V., and M.G.) inspected the MUSE data cubes  and the {\it HST} RGB images (F435W+F606W+F814W). Galaxies undergoing RPS  were identified by searching for extraplanar, unilateral tails/debris with emission lines in the MUSE data cubes and/or unilateral tails/ debris from {\it HST} images that were confirmed to belong to the galaxy from the MUSE redshifts. Passive galaxies were  classified based on the lack of emission lines; undisturbed galaxies are selected for having magnitude in the F606W band brighter than 23.5mag, being in the redshift range 0.2$<$$z$$<$0.55 and not showing peculiar features or asymmetries in their emission line morphology. 

\subsection{Sample selection} \label{sec:select}
In this analysis we aim at characterizing the star forming properties of galaxies, using the \Ha emission as tracer of star formation and OI, H$\beta $ and [NII] as proxy for selecting star forming regions (see below). We therefore apply a redshift cut to the sample ($z<0.42$) to ensure that all lines  are  within the MUSE wavelength coverage. 
We then consider all galaxies in the sample that are either ram pressure stripped or undisturbed and that have non negligible total SFR (they can be devoid of ionized gas in the center, but having a star forming tail), for a total of 29 ram pressure stripped and 26 control sample galaxies (26 in the clusters, 10 in the field).  We note that due to the small number statistics, we can not separate between field and cluster control sample galaxies and from now on we will treat them together for the sake of a stataistically-robust analysis, even though some environmental effects could still play a role  \citep[e.g.,][]{Vulcani2019GASPGalaxies, Franchetto2021}. Among the ram pressure stripped galaxies, 25 show ionized tails while four are truncated disks.

\begin{table*}
\caption{Properties of the galaxies analyzed in this paper: Galaxy name (ID), coordinates (RA, DEC), redshift (z), cluster membership (0= field galaxy, 1=cluster galaxy), Class (0= control sample, 1= RPS tail, 3 = RPS truncated disk), total stellar mass (Mass), SFR in the disk (SFR$_{disk}$) and total SFR (disk+tail, SFR$_{tot}$). The full table is available online. }
\centering 
\begin{tabular}{lllcccrrr}
\hline
  \multicolumn{1}{c}{ID} &
  \multicolumn{1}{c}{RA} &
  \multicolumn{1}{c}{DEC} &
  \multicolumn{1}{c}{z} &
  \multicolumn{1}{c}{Memb} &
  \multicolumn{1}{c}{Class} &
  \multicolumn{1}{c}{Mass} &
  \multicolumn{1}{c}{SFR$_{disk}$} &
  \multicolumn{1}{c}{SFR$_{tot}$} \\
  \multicolumn{1}{c}{} &
  \multicolumn{1}{c}{[h:m:s]} &
  \multicolumn{1}{c}{[deg]} &
  \multicolumn{1}{c}{} &
  \multicolumn{1}{c}{} &
  \multicolumn{1}{c}{} &
  \multicolumn{1}{c}{[10$^9$ $M_\sun$]} &
  \multicolumn{1}{c}{[$M_\sun/yr$]} &
    \multicolumn{1}{c}{[$M_\sun/yr$]} \\
\hline
  SMACS2031\_03 & 20:31:53.1 & -40:37:01.0 & 0.3177 & 1 & 0 & 0.6$\pm$0.3 & 0.037$\pm$0.007 & 0.051$\pm$0.01\\
  SMACS2031\_01 & 20:31:53.2 & -40:37:03.6 & 0.3523 & 1 & 1 & 46$\pm$13 & 0.07$\pm$0.01 & 0.19$\pm$0.04\\
  RXJ1347\_07 & 13:47:32.4 & -11:45:09.3 & 0.3468 & 0 & 0 & 2.7$\pm$0.9 & 0.12$\pm$0.02 & 0.12$\pm$0.02\\
  RXJ1347\_05 & 13:47:30.0 & -11:44:35.0 & 0.3084 & 0 & 0 & 5$\pm$2 & 0.13$\pm$0.03 & 0.13$\pm$0.03\\
  MACS1206\_23 & 12:06:16.4 & -08:47:43.6 & 0.3531 & 0 & 0 & 2$\pm$1 & 0.15$\pm$0.03 & 0.2$\pm$0.03\\
  MACS1206\_17 & 12:06:13.0 & -08:47:39.3 & 0.4106 & 1 & 0 & 18$\pm$7 & 1.8$\pm$0.4 & 2.0$\pm$0.4\\
  MACS1206\_15 & 12:06:11.0 & -08:48:22.3 & 0.4031 & 1 & 0 & 74$\pm$14 & 0.54$\pm$0.108 & 0.5$\pm$0.1\\
  MACS1206\_10 & 12:06:09.2 & -08:48:14.2 & 0.4224 & 1 & 1 & 1.0$\pm$0.6 & 0.047$\pm$0.009 & 0.09$\pm$0.02\\
  MACS1206\_05 & 12:06:13.0 & -08:47:39.4 & 0.411 & 1 & 0 & 19$\pm$8 & 1.9$\pm$0.4 & 2.0$\pm$0.4\\
  MACS1206\_04 & 12:06:11.8 & -08:47:49.4 & 0.4196 & 1 & 0 & 2.1$\pm$0.9 & 0.027$\pm$0.005 & 0.028$\pm$0.006\\
\hline\end{tabular}
\label{tab:gals}
\end{table*}

\section{Data analysis} \label{sec:analysis}

As discussed in \cite{Moretti2022}, we corrected the reduced datacube for extinction due to our Galaxy and subtracted the stellar-only component of each spectrum
derived with the spectrophotometric code \sinopsis \citep{Fritz2017}.  In addition to the best fit stellar-only model cube that is subtracted from the observed cube, \sinopsis provides for each MUSE spaxel stellar masses, luminosity-weighted and mass-weighted ages and star formation histories in four broad age bins. For more details we refer to \cite{Fritz2017},

We then derived emission line fluxes with associated errors using \highelf (M. Radovich et al. in prep).  We considered as reliable only spaxels with S/N(H$\alpha$,  H$\beta$, [OIII]5007, [NII])$>$2. { The data reach a surface brightness detection limit of  $\log(H\alpha [erg/s/cm^2/arcsec^2]) \sim -17.6 $ at the 2$\sigma$ confidence level.
}
\Ha luminosities corrected both for stellar absorption and for dust
extinction were used to compute SFRs, adopting the \cite{Kennicutt1998a}'s relation: $\rm SFR (M_{\odot}
\, yr^{-1}) = 4.6 \times 10^{-42} L_{\rm H\alpha} (erg \, s)$. 
The extinction was estimated from the Balmer decrement
assuming an intrinsic value of  $\rm H\alpha/H\beta = 2.86$ and the \cite{Cardelli1989} extinction law.  As the formal errors obtained by \highelf are negligible with respect to the uncertainties of the conversion factor from luminosities to SFR, we assume uncertainties on SFR to be 20\% of the values \citep{Kennicutt2009}. { We note that 95\% of the spaxels used in our analysis have a S/N$>$5 and the bulk of the spaxels have S/N$\sim$30, therefore their formal error is significantly smaller than the adopted 20\%.}

We employed the standard diagnostic diagram 
[OIII]5007/$\rm H\beta$ vs [NII]/$\rm H\alpha$ \citep{Baldwin1981} to separate the regions where the gas is ionized by star formation from regions powered by 
AGN or LINER emission. We adopted the division lines by \citet{Kewley2001, Kauffmann2003}. For the majority of the galaxies most of the \Ha emission is powered by photoionization (plots not shown).   Two galaxies (both undergoing ram pressure stripping) in the sample host an AGN: SMACS2031\_01  and A370\_06 \citep[see also][]{Moretti2022}. To compute SFRs, we considered only the spaxels for which ionized flux is powered by star formation  or  belong to the region of composite ionizing sources defined by \cite{Kauffmann2003} in the BPT diagram.\footnote{SFR values do not significantly change if we exclude Composite regions from the computation.}

We compute the total stellar mass and SFR of galaxies  by summing the values of all of the spaxels belonging to each galaxy and powered by star formation according to the BPT and we assume uncertainties on SFR to be 20\% of the values \citep{Kennicutt2009}, to take into account the uncertainties of the conversion factor from luminosities to SFR.
. 
To define the stellar disk region, we used the definition of galaxy boundaries developed by  \cite{Moretti2022}. The adopted procedure is based on the MUSE g-band reconstructed image. For each galaxy, we  identified the surface brightness of the sky background, by masking the galaxy itself and the neighbors, if any, and then  identified the stellar isophote corresponding to a surface brightness 3$\sigma$ above the measured sky background level. In the presence of neighbors, we adjusted the isophotes to remove their contributions. Since the resulting isophote can be quite jagged, we  fit an ellipse to the isophote. 
Everything inside of this contour represents the stellar disk, the rest constitutes the galaxy “tail''. Stellar masses were computed only within the ellipse defining the stellar disk, while for SFR we will also contrast disk and total (disk+tail) values.
By definition, control sample galaxies have negligible \Ha flux (therefore SFR) outside of the ellipse defining the stellar disk.

An extract of the properties of the galaxies used in this work is given in Tab.\ref{tab:gals}, while the full table is available online.
 
\section{Results} \label{sec:results}

\subsection{The global SFR-Mass relation}

\begin{figure*}
\centering
	\includegraphics[width=1\textwidth]{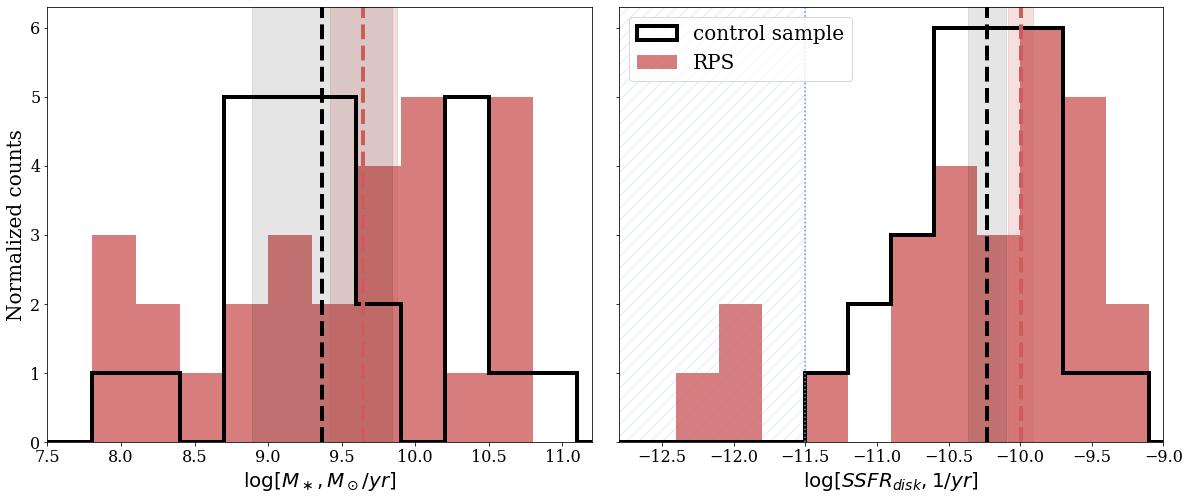}
	\caption{Left: Mass distribution (left) and disk SSFR (right) of the galaxies in the samples. Vertical lines and shaded areas show the median values with the uncertainties, computed as errors on the median. In the right panel, the blue shaded area indicates the region where SFR$_{disk}$/M$_\odot$$>$10$^{-11.5}$yr$^{-1}$.}
\label{fig:mass-ssfr_distr}
\end{figure*}

Figure \ref{fig:mass-ssfr_distr} shows the stellar mass (left) and the disk specific star formation rate (SSFR = SFR$_{disk}$/M$_\ast$, right) distributions, for ram pressure stripped and control sample galaxies. No clear differences emerge from the mass distributions: both the median values are compatible and the Kolmogorov-Smirnov (KS) test is unable to detect different parent distributions (statistic=0.20, pvalue=0.49). Considering the SSFR distributions, the median value of ram pressure stripped galaxies is shifted toward larger values, but again the KS test is not able to retrieve any significant difference (statistic=0.25, pvalue=0.28), most likely due to the small sample size. In the right panel of Fig. \ref{fig:mass-ssfr_distr} we also report the adopted separation between star forming and quiescent galaxies, that we set at SSFR$_{disk}$=10$^{-11.5}$yr$^{-1}$, as typically adopted in the literature \citep[e.g.][]{Salim2016}. Three galaxies fall below this threshold. In the following, even though we plot all galaxies, when performing the statistical analysis we will only consider galaxies that have a star forming disk.

\begin{figure*}
\centering
	\includegraphics[width=0.48\textwidth]{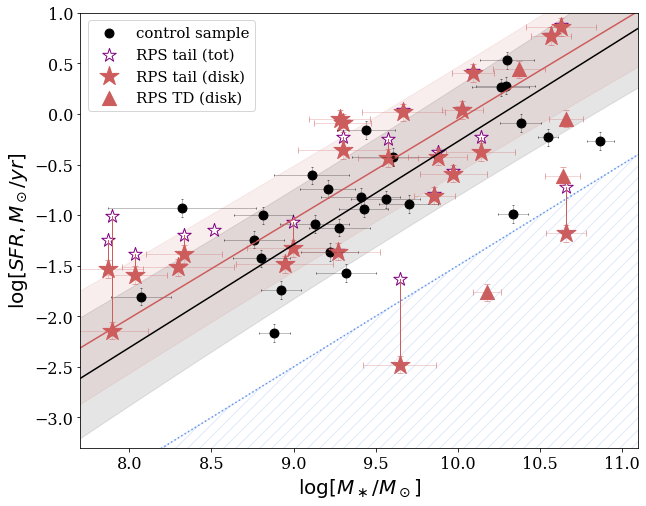}
	\includegraphics[width=0.48\textwidth]{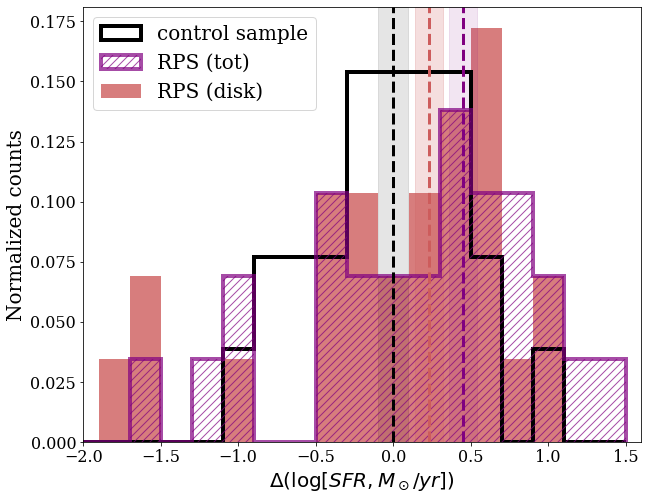}
	\caption{Left: SFR-mass relation for ram pressure stripped galaxies (stars) and control sample galaxies (black circles). For ram pressure stripped galaxies, filled red symbols refer to the SFR measured within the disk, empty purple symbols to the total SFR (disk+tail). Stars represent ram pressure stripped galaxies with ionized tails, triangles truncated disks. For each ram pressure stripped galaxy, the disk and total SFRs are connected by a line. If no empty star is plot, it means that the total and disk SFRs are comparable (hence the tail SFR is negligible). The black line and shaded grey regions show the best fit and 1$\sigma$ uncertainties for the control sample, while the red line and shaded red regions  show the best fit and 1$\sigma$ uncertainties for the ram pressure stripped sample, when the SFR$_{disk}$ is considered. The best fit obtained considering the total SFR is not shown, for the sake of clarity. Best fit values are reported in Tab.\ref{tab:best_fit}.  The dashed blue region indicates the plane where SSFR$_{disk}$$<$10$^{-11.5}$yr$^{-1}$.Right:  Distributions of the differences between the galaxy SFRs and their expected value according to the fit to the control sample, given their mass. Vertical lines and shaded areas show the mean values of the distributions along with uncertainties. Black and gray colors refer to the control sample, red (purple) colors show the values for the ram pressure stripped galaxies when the SFR in the disk (tot) is considered. }
\label{fig:mass-sfr_sample}
\end{figure*}

Figure \ref{fig:mass-sfr_sample} shows the main result of this paper: the SFR–\ma relation for galaxies undergoing RPS compared to that of  the control sample.
Among ram pressure stripped galaxies, those with tails populate the upper envelope of the control sample relation. This is true both when considering the  SFR$_{disk}$ values (filled red stars) and even more when considering the total (disk+tail) SFRs (empty purple stars). 
Truncated disk (filled red triangles), instead, tend to populate the lower envelope of the control sample relation, except for one case (MACS0416N\_01), which is on the main relation. 

Results are not driven by the galaxy inclination angles: no significant trends are observed when inclination is taken into account (plot not shown).

To statistically quantify the differences shown in Fig. \ref{fig:mass-sfr_sample} on a statistical ground, we fit a linear relation to each of  the datasets (considering RPS with tails and truncated disks together), using a least square fitting method that takes into account uncertainties on both axis. In all cases, we assume  the slope and intercept as free parameters. 
The best-fit values and the scatter of the relation are reported in Tab.\ref{tab:best_fit}. When considering  SFR$_{disk}$, ram pressure stripped and control sample galaxies are characterized by relations with very similar slopes, but the intercept is higher for the ram pressure stripped galaxies. The best fit relation instead flattens out when considering total SFRs. All relations are characterized by a similar scatter $\sigma$. Considering the upper 1-$\sigma$ envelope of the control sample best-fit relation, the fractions of galaxies above such relation are: 0.3$\pm$0.1 when considering SFR$_{disk}$, 0.4$\pm$0.1 when considering  SFR$_{tot}$ of ram pressure stripped galaxies, 0.08$^{+0.07}_{-0.04}$ when considering the control sample. If we repeat the fit fixing the slope for the ram pressure stripped galaxy sample at the value obtained for the control sample, we still obtain a difference in intercept of $\sim 0.3-0.4$ dex depending on the sample, in the sense that ram pressure stripped galaxies are systematically above,  even though differences are not statistically significant. 

This result indicates that galaxies feeling the effect of RPS show an enhancement of the SFR in the regions within the stellar disk, with respect to control sample galaxies of similar mass.
The same result is found { when the} SFR in the tails is considered. While in principle we should not compare disk+tails SFRs of the ram pressure stripped galaxies to the only disk SFRs of the control sample galaxies, we remind the reader that by definition control sample galaxies have no tails hence for them SFR$_{tail}$=0. To quantify the role of RPS though, it is important to quantify all the SFR induced by this mechanisms. Even though in the tails the ISM conditions are different with respect to those in the disks and there star formation is an indirect consequence of RPS, which only strips the gas that eventually will form new stars, we can not neglect this effect.

Differences between the various samples are better seen in the right panel of Fig. \ref{fig:mass-sfr_sample}, where the distribution of the difference between the SFR of each galaxy and the value derived from the control sample fit given the galaxy mass is shown. 
While a group of galaxies with reduced SFR is visible in all the samples, suggesting the presence of galaxies with a suppressed SFR, most of the stripping galaxies have a measured SFR higher than that expected given the fit, implying that their distribution is skewed toward higher values and is also broader. This result holds both when considering SFR$_{disk}$ and SFR$_{tot}$, but it is more evident in the second case. The KS test is able to retrieve significant differences between the control sample and the stripping sample when the SFR$_{tot}$ is considered (statistic=0.35, pvalue=0.05), while results are not significant when the disk SFR is considered (statistic=0.3, pvalue=0.19). While, by construction, the control mean value is centred at 0 (0$\pm$0.08 dex); mean values of the stripped galaxies distributions are the following: $\Delta \log (SFR_{disk}$)=  0.23$\pm$0.09 dex;  $\Delta \log (SFR_{tot}$)= 0.45$\pm$0.09 dex. Hence mean values are  different at 1.8(3.4)$\sigma$ level when considering the disk (total) SFRs. 

\begin{figure}
\centering
	\includegraphics[width=0.53\textwidth]{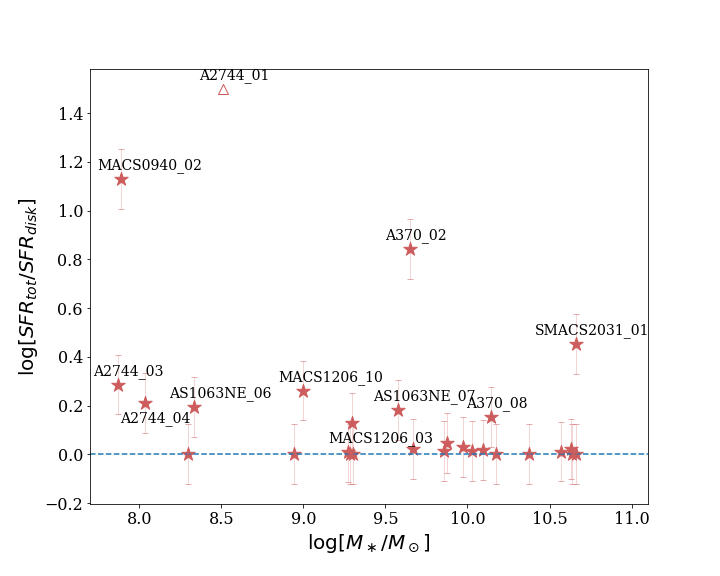}
	\caption{Logarithmic difference between the total and disk SFRs as a function of stellar mass,  for ram pressure stripped galaxies. Galaxies with a difference larger than 0.1 dex are labelled. A2744\_01, which has SDR$_{disk}=0$, is artificially located at y = 1.5. Errorbars are obtained summing  summing in quadrature errors on measurements. }
\label{fig:sfr_diff}
\end{figure}

Figure \ref{fig:sfr_diff} better quantifies the difference between the disk and total SFR for ram pressure stripped galaxies. In 11 out of 25 galaxies with tails, the tail SFR is non negligible and the total SFR is 1.25$\times$ higher than the disk SFR.  No trends with stellar mass nor with inclination (plot not shown) emerge.

\begin{table}
\caption{Best fit values for the SFR-mass relations shown in Fig.\ref{fig:mass-sfr_sample}. For the ram pressure stripped samples, best-fit values obtained fixing the slope to the control sample best-fit value are also given. }
\centering 
\begin{tabular}{lrrr}
\hline
  \multicolumn{1}{c}{Sample} &
  \multicolumn{1}{c}{slope} &
  \multicolumn{1}{c}{intercept}  & 
   \multicolumn{1}{c}{1-$\sigma$ scatter}  \\
\hline
control sample &1.02$\pm$0.15& -10.5$\pm$1.4 & 0.57\\
RPS (disk) &0.98$\pm$0.12& -9.9$\pm$1.1 & 0.54\\
RPS (tot) &0.86$\pm$0.15& -8.6$\pm$1.1 & 0.56\\
RPS (disk, fix) &1.02& -10.2$\pm$1.1 & 0.56\\
RPS (tot, fix) &1.02& -10.1$\pm$1.3 & 0.64\\
\hline\end{tabular}
\label{tab:best_fit}
\end{table}

\subsection{The spatially resolved SFR-Mass relation in the galaxy disks}
\begin{figure*}
\centering
	\includegraphics[width=0.95\textwidth]{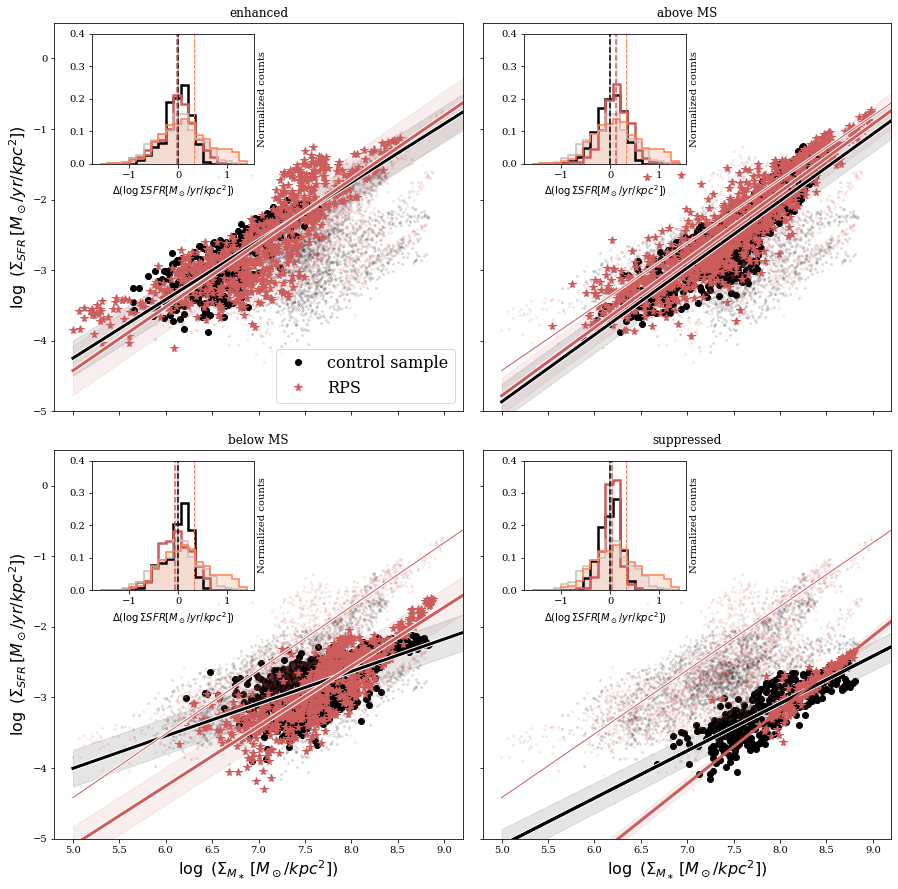}
	\caption{Spatially resolved SFR-mass relation for galaxies in different regions of the global SFR-mass plane Top left: enhanced region, top right: above MS region, bottom left: below MS region, bottom left: suppressed region. Small symbols show the distribution of all points, regardless of their position on the SFR mass plane and are reported in all panels. Black points are spaxels of control sample galaxies, black lines and grey shaded areas show the control sample best-fits along with the scatter, red stars are spaxels for RPS galaxies,  red lines and shaded areas the corresponding best-fits. In the suppressed panel we can not perform the fit to the RPS points because of their small sample statistics. In all panels, the best fit for the enhanced ram pressure stripped line is also reported in red, to ease comparisons. In the inset of each panel the distributions of the differences between the galaxy spatially resolved SFRs and their expected value according to the best-fit to the control sample, given their spatially resolved mass, considering the galaxies in that SFR-Mass region are reported. Vertical lines and shaded areas show the mean values of the distributions along with uncertainties. Black and gray colors refer to the control sample, red colors show the values for the ram pressure stripped galaxies. In each inset the same distributions and mean values considering all galaxies together are also reported (filled histogram and grey lines refer to the control sample, filled histogram and coral lines to the ram pressure stripped sample). }
\label{fig:mass-sfr_resolved}
\end{figure*}

We now wish to investigate more in detail the properties of galaxies located in different regions of the SFR-\ma plane, to look for additional pieces of evidence of the RPS effect on galaxy properties. To do so, adopting the control sample best fit and its scatter, we consider four different regions of Fig.\ref{fig:mass-sfr_sample}:
\begin{itemize}
    \item galaxies with $SFR_{disk}>SFR_{cs-fit}+ 1\sigma$ ({\it enhanced})
    \item galaxies with $SFR_{cs-fit}<SFR_{disk}<SFR_{cs-fit}+ 1\sigma$ ({\it above the main sequence\footnote{The term main sequence generally refers to an approximately linear relation between the SFR and \ma of star-forming galaxies \citep[e.g.][]{Noeske2007a,Noeske2007b}.}})
    \item galaxies with $SFR_{cs-fit}-1\sigma<SFR_{disk}<SFR_{cs-fit}$ ({\it below the main sequence})
    \item galaxies with $SFR_{disk}<SFR_{cs-fit}- 1\sigma$ 
    ({\it suppressed})
\end{itemize}
where $SFR_{cs-fit}$ is the SFR estimated from the fit for the control sample at any given mass and $1\sigma$ is the standard deviation of the fit.  The number of galaxies in each of the four groups is reported in Tab.\ref{tab:numbers}.
The different distribution of galaxies in the different regions is clear, with ram pressure stripped galaxies residing preferentially in the upper region of the plan, as already discussed. 

\begin{table}
\caption{Number of galaxies in the different regions of the SFR-\ma plane shown in Fig.\ref{fig:mass-sfr_sample}.}
\centering 
\begin{tabular}{lcc}
\hline
  \multicolumn{1}{c}{Region} &
  \multicolumn{1}{c}{RPS} &
  \multicolumn{1}{c}{control sample}   \\
\hline
enhanced & 9  & 2 \\
above MS & 7 & 12 \\
below MS & 8 & 9 \\
suppressed & 5 & 4\\
\hline\end{tabular}
\label{tab:numbers}
\end{table}

Figure \ref{fig:mass-sfr_resolved} compares the spatially resolved SFR-\ma relation (\Ssfr-\Sm) for galaxies in the different regions of the SFR-\ma plane, affected or not by RPS. For fair comparisons, only spaxels within the galaxy disks are considered. Values are deprojected by considering the axis ratio of the ellipse defining the disk. In total, 3271 points are plotted for the control sample galaxies, 4368 points are  plotted for the ram pressure stripped galaxies.

Overall, a  correlation  exists between the \Ssfr and \Sm  on a few kpc scales, similarly to what is seen in the Local Universe \citep{Ellison2018, Vulcani2019GASPGalaxies, Brown2023}. While the  plot considering the different zones together is quite scattered (small colored symbols plotted in the background of all plots) -- indicating a large galaxy-by-galaxy variation in star forming properties on local scales, similarly to what seen in \cite{Vulcani2019GASPGalaxies, Vulcani2020b} -- cleaner trends emerge when  galaxies are divided in the four groups. In each panel, galaxies cover  similar regions of the plane, notwithstanding the RPS is ongoing or not. Nonetheless, distinct sub-relations appear and these correspond to different galaxies: to some extent, each object spans a different locus of the \Ssfr-\Sm plane. 
In addition, the relative distribution of points in the different panels depends on the sample: in the case of the control sample, 13.8$\pm$0.6\% of the data points are in the enhanced region, 37.2$\pm$0.8\% in the above the main sequence region, 32.3$\pm$0.8\% in the below the main sequence region and 16.7$\pm$0.6\% in the suppressed region.  The corresponding fractions for the ram pressure stripped samples are as follows: 21.7$\pm$0.6\%, 35.3$\pm$0.7\%, 26.6$\pm$0.6\% and 16.4$\pm$0.5\%. 
The distribution of the spaxels seems not to depend on the galaxy total stellar to be independent on  inclination and galactocentric distance of the spaxels (plots not shown). \cite{Vulcani2019GASPGalaxies} did find that relations are driven by the presence of bright star forming clumps spread across the galaxy disks, but the poor spatial resolution of MUSE data at $z\sim 0.35$, which is $\sim 5$kpc/\arcsec prevents us from understanding if this is the case also at this epoch. 

\begin{table*}
\caption{Best fit values for the \Ssfr-\Sm relations shown in Fig.\ref{fig:mass-sfr_resolved}.
}
\centering 
\begin{tabular}{lrrr|rrr}
\hline
  \multicolumn{1}{c}{Sample} &
  \multicolumn{3}{c|}{RPS} &
  \multicolumn{3}{c}{control sample} \\
  & 
  \multicolumn{1}{c}{slope} &
  \multicolumn{1}{c}{intercept}  & 
   \multicolumn{1}{c|}{1-$\sigma$ scatter}  &
     \multicolumn{1}{c}{slope} &
  \multicolumn{1}{c}{intercept}  & 
   \multicolumn{1}{c}{1-$\sigma$ scatter}  \\

\hline
enhanced & 0.903$\pm$0.022 & -8.93$\pm$0.16 & 0.35 & 0.829$\pm$0.029 & -8.39$\pm$0.19 & 0.25 \\
above MS & 0.962$\pm$0.011 & -9.59$\pm$0.09 &0.25 & 0.949$\pm$0.016 & -9.61$\pm$0.12 & 0.25\\
below MS & 0.843$\pm$0.021 & -9.32$\pm$0.16 &0.27& 0.458$\pm$0.021 & -6.29$\pm$0.16 & 0.26\\
suppressed & 1.046$\pm$0.027 & -11.55$\pm$0.23 & 0.09 & 0.67$\pm$0.04 & -8.45$\pm$0.30 & 0.21\\
\hline\end{tabular}
\label{tab:best_fit_spat}
\end{table*}

Considering each panel of Fig.\ref{fig:mass-sfr_resolved} separately, the best-fit parameters  of linear relation fitted  to the different samples (Tab.\ref{tab:best_fit_spat}) show that all trends for the ram pressure stripped samples are slightly steeper than for the control sample relation, suggesting enhanced \Ssfr at high \Sm values.\footnote{{ Values of the fits do not significantly change when we adopt a more strict S/N cut (S/N$>$5) and all results hold.}} Overall, comparing the best-fit parameters for a given sample across the different panels, we find a progressive flattening of the relations as we move down along the global SFR-\ma plane. 
In the insets,  the distributions of the differences between the galaxy \Ssfr and their expected value according to the best-fit to the control sample in that region, given their \Sm are reported in red for ram pressure stripped galaxies and in black for the control sample, along with mean values. Distributions of the ram pressure stripped samples are systematically shifted towards larger values, except for the region below the main sequence. Both the KS test and the comparison of the mean values support this finding: the KS test run pairwise recovers significant differences (p-value $<$0.0005) in all cases. 
In each inset, we also report for comparison the distributions  of the differences between the galaxy \Ssfr and their expected value according to the best-fit to the full control sample, when galaxies are considered together, using filled histograms. In this case,  differences are even more enhanced: the ram pressure stripped sample is shifted towards larger $\Delta$ values at ($>10\sigma$). This result is mainly driven by the larger density of points in the enhanced region of ram pressure stripped galaxies.

To summarize, similarly to what found in \cite{Vulcani2020b} in the Local Universe, galaxies above the global SFR-\ma relation are also found above the spatially resolved \Ssfr-\Sm  relation, which means that in those galaxy regions that still sustain star formation, ram pressure stripped galaxies form stars at a higher rate per unit of galaxy area than undisturbed galaxies. Taking all galaxies together, ram pressure stripped galaxies have an enhancement of \Ssfr at any given \Sm. This enhancement is more pronounced at higher \Sm values (\Sm$>10^{7.5}$\msk), in agreement to  low-z results  \citep{Vulcani2020b}. 

\section{Summary and discussion} \label{sec:summary}

In the Local Universe, multiple observations point to an enhancement of the SFR in cluster galaxies undergoing ram pressure stripping \cite[e.g][]{Poggianti2016JELLYFISHREDSHIFT, Vulcani2018_L, Vulcani2019GASPGalaxies, RobertsParker2020}. This result is supported by theoretical works in some cases \citep[e.g.][]{Kronberger2008, TonnesenBryan2012, Roediger2014, Troncoso2020}, even though some other studies do not predict such trend \citep{2023arXiv230409199G}.

In this paper, we investigate for the first time  the role of ram pressure stripping on the star forming properties of galaxies beyond the Local Universe on a statistical ground. \cite{Moretti2022} and A. Moretti et al.  (in prep.) have assembled the  largest sample with available spatially resolved spectroscopy and reliable classification 
of ram pressure stripped galaxies at $z\sim 0.35$, exploiting  MUSE observations gathered by the MUSE GTO \citep{Richard2021}. They also assembled a coeval control sample of cluster and field galaxies not affected by ram pressure stripping, allowing us to perform systematic comparisons between the different samples. 

We measure SFRs from \Ha emission for 29 ram pressure stripped galaxies  (25 with ionized tails, 4 truncated disks) and 26 control sample galaxies. For the ram pressure stripped sample, we consider both the SFR within the galaxy disk and the SFR generated in the stripped tails. 
Both considering only the disk SFR and the disk+tail SFR, we find that ram pressure stripped galaxies are characterized by a much steeper SFR-\ma relation than control sample galaxies, confirming the results of \cite{Lee2022}, based on a sample of 5 ram pressure stripped galaxies. The best fit parameters are different at $\sim 2\sigma$ level when the disk SFR is considered. Considering the difference between the observed SFR and that expected given the control sample best fit and the mass of each galaxy, we find that galaxies in the ram pressure stripped sample are characterized by a clear enhancement in SFR with respect to their control sample counterparts. The difference between the mean control  sample value and the ram pressure stripped sample is 0.3 dex when the disk SFR is considered (1.8$\sigma$ significance) and 0.44 dex when the total SFR is considered (3.4$\sigma$ significance). 

We note that this difference might also be just a lower limit. Even though we corrected SFRs for the presence of dust by using the Balmer decrement, a large amount of dust enshrouded star formation could characterize ram pressure stripped galaxies \citep{Rawle2014}. \cite{Vulcani2023a} analyzed the JWST/NIRSpec spectrum for one of the galaxies in our sample (A2477\_10) showing it is dominated by a strong PAH in emission at 3.3$\mu m$, a line typically considered as tracer of dust obscured star formation \citep{Peeters2004, Brandl2006}. More in general, \cite{Vulcani2023a}, analyzing JWST/NIRCam data of all galaxies in Abell 2744, found that the ram pressure stripped ones (which also enter our sample) are characterized by extremely red F200W-F444W colors, hypothesizing that this class of objects is indeed characterized by dust obscured star formation, invisible at optical wavelengths. Larger coverage of NIRSpec data, mapping the entire galaxies, is needed to securely establish the connection between red IR colors and the presence of the PAH.

Thanks to the MUSE data, we can also investigate the spatially resolved SFR-Mass relation, on a spatial scale of approximately 5 kpc/\arcsec. Overall, the ram pressure stripped sample is characterized by a steeper \Ssfr-\Sm relation and the SFR enhancement is retrieved also on local scale. Controlling for the location of galaxies on the global SFR-Mass plane, differences are reduced, but overall still present. 

To conclude, we have retrieved most of the trends observed in ram pressure stripped galaxies in the Local Universe \citep{Vulcani2018_L, Vulcani2019GASPGalaxies, Vulcani2020b, Brown2023}, showing how this mechanism affects the star forming properties of galaxies at $z\sim0.35$. Similarly, Khormal et al. (submitted) has shown that RPS has also effects similar to those in the Local Universe for as far as the gas abundances in concerned, with ram pressure stripped tails being characterized by low metallicities due to the interaction between the ICM and ISM \citep{Franchetto2021_tails}. What remains to be addressed is  the effective role of ram pressure in the quenching and morphological transformations that are known to have involved a large fraction of today’s clusters galaxies since $z\sim 1$ \citep{Pallero2019, Pintos2019} and  when ram pressure starts playing its significant role.

\begin{acknowledgements}
We thank the referee for their comments. This project has received funding from the European Research Council (ERC) under the Horizon 2020 research and innovation programme (grant agreement N. 833824). 
\end{acknowledgements}

\bibliography{references.bib}{}
\bibliographystyle{aa}



\end{document}